Improved Photocatalytic Performance via Air-Plasma Modification of Titanium Dioxide: Insights from Experimental and Simulation Investigation

Verbesserte photokatalytische Leistung durch Luftplasma-Modifikation von Titandioxid

C. Ugwumadu[1,*], I. Olaniyan[2,3], H. K. Jeong[4], and D. A. Drabold[1,*]

[1]Department of Physics and Astronomy, Nanoscale and Quantum Phenomena Institute (NQPI), Ohio University, Athens, Ohio 45701, USA.

[2]Helmholtz-Zentrum, Berlin 14109, Germany

[3]Department of Physical Chemistry, Freie Universität Berlin, Berlin 14195, Germany

[4]Department of Materials-Energy Science and Engineering, Institute of Industry and Technique Daegu University, Gyeongsan 38453. Republic of Korea

*Corresponding authors E-mail: cu884120@ohio.edu; drabold@ohio.edu

Abstract:

Commercial titanium dioxide is successfully plasma-treated under ambient conditions for different time periods, leading to reduced crystallite size and the creation of oxygen vacancies. Density functional theory-based calculations reveal the emergence of additional localized states close to the conduction band, primarily associated with under-coordinated titanium atoms in non-stoichiometric titanium-oxide systems. The plasma-treated samples exhibit improved photocatalytic performance in the degradation of methylene blue compared to untreated samples. Moreover, the 4-hour plasma-treated photocatalyst demonstrates commendable stability and reusability. This work highlights the potential of cost-effective plasma treatment as a simple modification technique to significantly enhance the photocatalytic capabilities of titanium dioxide.

Abstrakt:

Kommerzielles Titandioxid wird unter Umgebungsbedingungen über verschiedene Zeiträume erfolgreich plasmabehandelt, was zu einer Verringerung der Kristallitegröße und zur Bildung von Sauerstofflücken führt. Berechnungen basierend auf der Dichtefunktionaltheorie zeigen das Auftreten zusätzlicher lokalisierter Zustände in der Nähe des Leitungsbandes, die in erster Linie mit unterkoordinierten Titanatomen in nicht-stöchiometrischen Titanoxid-Systemen assoziiert werden. Im Vergleich zu den unbehandelten Proben zeigen die plasmabehandelten eine verbesserte photokatalytische Leistung beim Abbau von Methylenblau. Darüber hinaus weist der 4 Stunden plasmabehandelte Photokatalysator eine beachtliche Stabilität und Wiederverwendbarkeit auf. Diese Arbeit veranschaulicht das Potenzial einer kostengünstigen Plasmabehandlung als einfache Modifikationstechnik zur deutlichen Verbesserung der photokatalytischen Fähigkeiten von Titandioxid.

Schlüsselwörter: *titandioxid, plasmabehandlung, photokatalyse, molekulare simulation, ultraviolettes Licht*

# 1 Introduction

Industrial activities, such as leather, paper, and textile production, generate hazardous waste containing the cationic dye, methylene blue (MB), which poses significant health risks and requires effective removal from industrial effluents [1, 2]. Titanium dioxide ($TiO_2$) and its composites have emerged as highly versatile compounds with applications in hydrogen evolution reaction [3] and sensor devices [4], but their most crucial application lies in the photocatalytic degradation of industrial dyes [5, 6]. However, titanium dioxide faces limitations such as low surface area, wide bandgap, and rapid charge recombination, hampering its photocatalytic efficiency.

To overcome those limitations, this study explores the improved photocatalytic activity of plasma-modified titanium dioxide under ambient conditions for various time intervals (1-4 hours). The investigations include changes in its morphology, chemical speciation, and charge-transfer resistance, along with the evaluation of the composites' photocatalytic performance in methylene blue degradation. Furthermore, the study employs density functional theory using the Vienna *ab initio* simulation package (VASP) [7] to study the impact of oxygen vacancies in sub-stoichiometric titanium suboxide models ($TiO_{2-x}$, where $x < 2$) on the electronic density of states. These findings highlight the potential of air-plasma treatment as a cost-effective and promising approach to tailor titanium dioxide properties for efficient industrial dye degradation and environmental remediation.

# 2 Materials and Methods

## 2.1 Sample Preparation

50 mg of titanium dioxide (99.8%, CAS No: 1317-70-0, Sigma-Aldrich) was uniformly spread in a flat-bottom tube ($35 \times 12$ mm soda lime glass) and wetted with 200 $\mu$l of deionized water to ensure surface uniformity. The plasma treatment was conducted at atmospheric pressure using a discharge potential of 15 kV and a frequency of 25 kHz. The anode material was a stainless-steel needle that was positioned 20 mm apart from a copper tape cathode as described elsewhere [8, 9, 10, 11].

The prepared samples were divided into five groups based on the duration of plasma treatment (ranging from 0 to 4

hours). To minimize statistical error, three samples were prepared in each group for subsequent tests, resulting in a total of 15 samples. The plasma-treated samples were labeled as P$\xi$h (where $\xi$ = 1, 2, 3, 4), while three precursors (untreated samples) served as controls and were labeled as P0h. After plasma treatment, the samples were washed and filtered using deionized water followed by drying in a vacuum oven at 60 °C for 24 hours. Methylene blue dye, purchased from Sigma-Aldrich, was used without any treatment to assess the photocatalytic degradation capability of the samples.

## 2.2 Structural, Physiochemical and Electrochemical Analysis

The surface morphology and elemental analysis were examined using a field emission scanning electron microscope with an integrated energy-dispersive x-ray spectrometer. The structure and crystallite size of the prepared samples was analyzed using an x-ray diffractometer with a copper $K_\alpha$ radiation source. The chemical configuration was investigated using x-ray photoelectron spectroscopy with an aluminum $K_\alpha$ x-ray source, and a pass energy of 20 eV.

Photo-electrochemical characterizations were performed using a three-electrode system with one mole of sulfuric acid used as the electrolyte. The Bio-logic (SP-150) EC-Lab instrument was used for electrochemical performance investigations. The working electrode was a glassy carbon electrode, with silver / silver-chloride as the reference electrode and platinum wire as the counter electrode. For analysis, a dispersion of the sample in isopropyl alcohol was drop-casted on the working electrode. Additionally, Electrochemical impedance spectroscopy was conducted in the frequency range of 100 MHz to 500 kHz for photo-electrochemical characterization.

The photodegradation experiments were conducted at room temperature using methylene blue solution with a concentration of $1.6 \times 10^{-4}$ M (10 mg in 200 mL of deionized water). 40 mg of the samples were added to this solution in a 35 × 12 mm flat bottom glass tube. The mixture was magnetically stirred in the dark for 30 minutes to establish absorption/desorption equilibrium. At specific time intervals during the irradiation, 3 mL of the solution was extracted and centrifuged at 7,000 rpm for 30 minutes to remove any remaining photocatalyst. The dye content in the centrifuged solution was analyzed using an ultraviolet–visible spectroscopy spectrometer. Additionally, the photocurrent responses of the samples were measured at a constant potential of +1.0 V with on-off intervals of ultraviolet irradiation lasting 30 seconds.

### 2.3 Molecular Simulation

Three models of stoichiometric titanium dioxide (TiO$_2$) were generated using the quench-from-melt scheme, which has been successfully utilized in building models of titanium-oxides [12,13,14]. Using this approach, 192 atoms, consisting of 64 and 128 titanium and oxygen atoms respectively, were randomly distributed in a cubic box to achieve a density of 3.8 g/cm$^3$. To simulate melt, the models were annealed at 2500 K for 9 ps, After the first 2 ps, 3 liquid titanium dioxide configurations were selected at 2 ps intervals till the end of the annealing simulation. Each model was then independently cooled to 2200 K over 5 ps and equilibrated for another 5 ps. The models were further cooled to 1200 K at a rate of 80 K/ps and then equilibrated at this temperature for 2 ps. After this, they were taken to 300 K at a rate of 75 K/ps followed by conjugate gradient relaxation to an energy-minimum configuration.

To model oxygen-vacancies in non-stoichiometric titanium suboxide, depicted as TiO$_{2-x}$ (where $x < 2$), oxygen atoms were randomly removed from the stoichiometric models. The resulting systems were then heated at 550 K for 2 ps, and then cooled to 300 K over 2.5 ps. Subsequently, the models were relaxed to an energy-minimum configuration via conjugate gradient relaxation. For this study, we present results for models with 1, 5, and 10 oxygen-vacancies, enabling an investigation of the impact of varying vacancy concentrations on titanium-oxide systems.

The calculations were implemented within the Vienna *ab initio* simulation package (VASP) [7]. The Perdew-Burke-Ernzerhof (PBE) generalized gradient approximation (GGA) was used for the exchange-correlation functional in the simulations [15]. A plane-wave basis set with a kinetic energy of 450 eV was used to expand the electronic wave functions. For static calculations, a larger cutoff of 540 eV was utilized. The simulations were performed under periodic boundary conditions using a single Γ **k**-point.

# 3 Results and Discussion

The x-ray diffraction peaks for the untreated sample (P0h) were observed at 25.1°, 36.7°, 37.6°, 38.3°, 47.9°, 53.7°, 54.9°, and 61.9°. These peaks align close to the (101), (103), (004), (112), (200), (105), (211), and (204) crystallographic planes of anatase titanium dioxide (JCPDS no. 00-021-1272), Figure 1. While the plasma-treated samples exhibited similar diffraction peaks, their intensities decreased with increasing treatment duration. Notably, for P4h, the (103) and (112) peaks nearly disappeared. Additionally, the prominent peak at 25.1° slightly shifted to 25.08°, 24.72°, 24.56°, and 24.58° in P1h, P2h, P3h, and P4h, respectively. These shifts and intensity reductions in the diffraction peaks could have resulted from several factors, including structural modifications, lattice defects, or the presence of impurities due to the plasma process.

To investigate the underlying cause of the reduced peak intensity, we utilized the modified Scherrer formula to obtain the average crystallite size of the samples based on the width of the x-ray diffraction peaks [16]. We start by taking the logarithm of the basic Scherrer formula [17]:

$$ln\beta = ln\frac{K \cdot \lambda}{D} + ln\frac{1}{cos(\theta)} \qquad (1)$$

where $D$ is the average crystallite size, $K$ is the shape factor (typically taken as 0.9), $\lambda$ is the x-ray wavelength, $\beta$ is the full width at half-maximum (FWHM) of the diffraction peaks for $2\theta$ values corresponding to the (101), (105) and (204) planes. The single-valued crystallite size ($D$), which decreased with increasing plasma treatment duration, was obtained from the exponent of the intercept ($e^{ln\frac{K \cdot \lambda}{D}}$) in the linear plot of $ln\beta$ vs $ln(1/cos\theta)$, second column in Table 1. Additionally, scanning electron microscope images of the sample's surface morphology indicates that the samples exhibited a finer structure as the plasma treatment time increased, Figure 2. Particularly, P4h displayed a smooth and refined texture compared to the rough structure of P0h - a consequence of the reduced crystallite size.

The elemental composition of the samples from the energy dispersive spectroscopy measurements indicates that the plasma treatment significantly reduced the relative atomic weight percentage of oxygen in the treated samples (Table 1, third and fourth). This reduction in oxygen content prompted further investigations into the chemical state of the samples, carried out using x-ray photoelectron spectroscopy measurements. The results revealed a remarkable shift in the peaks corresponding to titanium-2p states, Figure 3. In the pristine sample (P0h), distinct titanium 2p doublet peaks were observed at 459.1 eV and 464.8 eV, representing the $Ti^{4+}$ $2p_{3/2}$ and $Ti^{4+}$ $2p_{1/2}$ states, respectively. However, in P4h, these doublet states were shifted to higher energies, specifically at 459.5 eV and 465.4 eV, respectively. Interestingly, a deconvolution analysis of the P4h spectrum revealed the presence of an additional state at 460.1 eV, corresponding to the $Ti^{3+}$ $2p_{3/2}$ state. The observed shift in energy and the emergence of the $Ti^{3+}$ state in the plasma-treated sample result from oxygen vacancies within the system. These vacancies create donor states within the bandgap of titanium dioxide [18].

Comparing the electronic density of states (EDoS) calculated from density functional theory for the stoichiometric titanium-oxide ($TiO_2$) and non-stoichiometric titanium-oxide ($TiO_{2-x}$) models, showed that the oxygen-vacancies in the non-stoichiometric models create new Kohn-Sham states in the HOMO-LUMO gap region, Figure 4. The energy proximity of these new Kohn-Sham states to the conduction bands suggest that they are shallow states, with more states created with increasing oxygen vacancies in the models. It is important to note that a single Γ **k**-point in the supercell was used for this calculation, and as such, the subsequent discussions are approximations based on this specific Brillouin zone sampling. The HOMO-LUMO gap of the stoichiometric titanium dioxide is approximately 2 eV, consistent with previous reports for anatase phase of titanium dioxide [19, 20].

The creation of the gap states is accompanied by a shift of the Fermi level further away from the valence band as the number of vacancies increases, making the non-stoichiometric titanium dioxide more p-type. These shallow states shift the position of the Fermi level and alter the redox potential of the photocatalyst That facilitate specific reaction pathways and promote the activation of certain species, enhancing the photocatalytic activity and selectivity of titanium dioxide [21]. Furthermore, the extent of localization of the Kohn-Sham states ($\phi$) was quantified using the electronic inverse participation ratio (EIPR) as defined by Equation 2 [22]:

$$I(\phi_n) = \frac{\sum_i |a_n^i|^4}{\left(\sum_i |a_n^i|^2\right)^2} \qquad (2)$$

where $a_n^i$ denotes the contribution to the eigenvector ($\phi_n$) from the $i^{th}$ atomic orbital. The EIPR with high (low)

values indicate localized (extended) states, depicted by the gray lines in Figure 4. The specific atoms responsible for states within +2 eV of the Fermi level in the 5 and 10 oxygen-vacancy models are predominantly localized on the under-coordinated titanium atoms (refer to Figure 5). The presence of these shallow states introduces additional energy levels within the HOMO-LUMO gap in titanium dioxide, which serve as effective traps for photogenerated electrons. This trapping mechanism significantly reduces rapid recombination with holes [6, 21]. By inducing electron-hole recombination, the shallow trap states promote efficient charge carrier separation, enabling photogenerated electrons to actively engage in redox reactions on the surface of the photocatalyst. The x-ray photoelectron spectroscopy data showing the presence of $Ti^{3+}$ states, and the localized shallow trap states obtained from density functional theory calculations, are supported by the photo-degradation of methylene blue under ultraviolet irradiation, Figure 6. The percentage of photo-degradation ($D$) of the dye is determined based on the initial dye concentration ($C_0$) and the concentration at the observation time t ($C_t$) as:

$$D = \left(1 - \frac{C_0}{C_t}\right) \times 100\% \qquad (3)$$

The optimal point of degradation for methylene blue was determined by its convergence to a specific value. In the case of samples P0h and P1h, this convergence occurred after 240 minutes, resulting in an approximate 74% degradation of methylene blue. Conversely, sample P4h achieved convergence after just 180 minutes, with a mere 7% of the dye remaining in the solution. Detailed snapshots of the methylene blue degradation process using P4h, and the corresponding UV-Vis spectra can be found in Figure S1a and S1b in the supplementary material. These findings highlight the superior photocatalytic performance of sample P4h, which exhibits faster, and more efficient degradation of methylene blue compared to the other samples. It is worth noting that all samples displayed significant degradation of the methylene blue dye, Figure S1c.

To quantitatively evaluate the photocatalytic activity of the catalysts, the reaction rate constants ($k$) were determined by first-order approximation of the Langmuir-Hinshelwood kinetic equation given as [23]:

$$-ln\left(\frac{C_0}{C_t}\right) = kKt \approx \kappa t \qquad (4)$$

$k$ and $K$ are the degradation rate constant and equilibrium constant for the adsorption of methylene blue by the catalyst respectively. $\kappa$ is the apparent rate constant [24]. The apparent rate constants of 0.018, 0.018, 0.019, 0.022, and 0.025 min$^{-1}$ were obtained for P0h, P1h, P2h, P3h, and P4h, respectively. To attain these results, only the data from 30 to 210 minutes were utilized, Figure S1d. The increased apparent rate results from the shallow trap states in the plasma-treated samples that facilitates the separation of photo-generated electron-hole pairs. This allows more sites to become available for the adsorption of water molecules to produce hydroxyl radicals (•OH), and thus increase the degradation rate.

Based on these results, our proposed photo-degradation mechanism of methylene blue using the plasma-treated samples involves an initial step wherein oxygen vacancies are formed within the titanium dioxide lattice. This leads to the formation of a modified material, denoted as TiO$_{2-x}$, which signifies the presence of oxygen vacancies:

$$TiO_2 + \text{heat} \rightarrow TiO_{2-x} \qquad (5)$$

Furthermore, in the presence of these oxygen vacancies, TiO$_{2-x}$ has a high affinity to adsorb oxygen molecules (O$_2$) onto its surface [25]:

$$TiO_{2-x} + O_2 \rightarrow TiO_{2-x}(O_2) \qquad (6)$$

Under the influence of ultraviolet irradiation ($h\nu$), the adsorbed oxygen molecules on TiO$_{2-x}$ becomes energized to an excited state (TiO$_{2-x}$(O$_2$)$^*$):

$$TiO_{2-x} + h\nu \rightarrow TiO_{2-x}(O_2)^* \qquad (7)$$

The presence of Ti$^{3+}$ species facilitates the transfer of electrons to the adsorbed oxygen, resulting in the formation of super-oxide radicals (O$^-$) and the conversion of Ti$^{3+}$ to Ti$^{4+}$:

$$TiO_{2-x}(O_2)^* + Ti^{3+} \rightarrow TiO_{2-x} + O_2^- + Ti^{4+} \qquad (8)$$

Finally, the super-oxide radicals proceed to undergo a reaction with water, yielding hydroxyl radicals (•OH). This crucial reaction is notably enhanced in the plasma-treated samples, which exhibit improved hydrophilicity—a prerequisite for hydroxyl radical generation, as detailed in Section S2 of the supplementary material. These hydroxyl radicals then interact with the methylene blue dye, catalyzing its degradation into a range of intermediate products and, ultimately, achieving complete mineralization:

$$O_2^- + H_2O \rightarrow \bullet OH + \bullet OH \qquad (9)$$

$$\text{methylene blue} + \bullet \text{OH} \rightarrow \text{Degradation Products} \qquad (10)$$

Electrochemical impedance spectroscopy was employed to investigate the surface charge transfer resistance of the samples, providing valuable insights into their electrochemical behavior. The semicircular Nyquist plot corresponds to the charge transfer resistance at the electrode interface, Figure 7. To analyze this impedance behavior, an equivalent circuit was utilized. The circuit comprises several components: $R_B$ and $R_{CT}$ represent the bulk resistance of the electrolyte and the charge transfer resistance respectively. W is Warburg impedance and CPE corresponds to the constant phase element (see inset in Figure 7). The Nyquist plot clearly illustrates that P4h exhibits the smallest semicircle, indicating the most rapid interfacial charge transfer and efficient separation of photo-generated charges. The calculated charge transfer resistances for P0h, P3h, and P4h were found to be 19.8 Ω, 17.7 Ω, and 16.9 Ω, respectively. These findings suggest that time-dependent air plasma treatment of titanium dioxide effectively retards the recombination of photo-generated charge carriers and facilitates interfacial charge transfer.

To gain deeper insights into the recombination rate of the composite, the photocurrent responses of the samples were also examined. The initial photocurrent responses of the samples upon toggling the ultraviolet on and off showed sharp anodic spikes, indicating the separation of electron/hole pairs at the catalyst/electrolyte interface, Figure 8. Subsequently, a time-dependent decrease in photocurrent occurs until reaching a steady-state current. This decline signifies the occurrence of electron-hole recombination. At the equilibrium current, a delicate balance is achieved between the competing processes of separation and recombination of electron-hole pairs. Notably, P4h exhibits the highest photocurrent, displaying a pronounced anodic peak. This suggests a delayed recombination process, leading to a more efficient separation of electron/hole pairs. Furthermore, the gradual decay of photocurrent to zero upon light deactivation indicates the release of charge carriers from shallow trap states. This observation implies a higher concentration of oxygen-vacancies in P4h resulting from extended plasma treatment.

The reusability of the P4h photocatalyst for methylene blue degradation was examined through the analysis of transient photocurrent response. The P4h photocatalyst was subjected to centrifugation and reused in four consecutive cycles for methylene blue degradation under ultraviolet irradiation. The percentage degradation in cycles 1, 2, 3, and 4 was 93.3%, 91.9%, 90.2%, and 87.8%, respectively. Even after the fourth cycle, the efficiency of the photocatalyst remained above 80%. This demonstrates the excellent stability, sustainability, and reusability of the air plasma-treated titanium dioxide photocatalyst.

## 4 Conclusion

This study presents the successful preparation of plasma-treated titanium dioxide under ambient conditions for extended durations, leading to a reduced crystallite size in titanium dioxide. The formation of shallow trap states is also evident, supported by calculations based on density functional theory, which reveals that non-stoichiometric models of titanium dioxide exhibit additional localized states in the HOMO-LUMO gap near the conduction band, primarily associated with titanium atoms. Furthermore, the utilization of plasma-treated materials in the degradation of methylene blue demonstrates enhanced performance compared to untreated samples. Particularly noteworthy is the 4-hour plasma-treated photocatalyst, exhibiting commendable stability and reusability.

Overall, this work highlights a simple and cost-effective modification of titanium dioxide through air plasma treatment as a promising candidate for advanced photocatalytic applications, offering an efficient and sustainable approach to address environmental pollution challenges. Future studies can further explore the optimization of plasma treatment parameters and investigate the underlying mechanisms to harness the full potential of this modified photocatalyst.

**Acknowledgements**

The authors acknowledge the National Science Foundation (NSF) for computational support through XSEDE (Grant No. ACI-1548562; allocation no. DMR-190008P) and ACCESS (Grant No. 2138259, 2138286, 2138307, 2137603, and 2138296.; allocation no. phy230007p). Basic Science Research Program through the National Research Foundation of Korea (NRF-2020R1I1A3A04037469) also supported this work. C. U. thanks Anna-Theresa Kirchtag for proofreading the manuscript.

## Author Contribution

C. Ugwumadu was involved in conceptualization, data curation, formal analysis, funding acquisition, investigation, methodology, writing – original draft, review & editing. I. Olaniyan was involved in validation, visualization and writing – review & editing. H. K. Jeong and D. A. Drabold were involved in conceptualization, supervision, funding acquisition, and review & editing.

## Conflict of Interest

The authors declare no financial or commercial conflict of interest.

## Data Availability Statement

The data that support the findings of this study are available from the corresponding author upon reasonable request.

## References

[1] Z. Mulushewa, W. T. Dinbore, Y. Ayele, *Environmental Analysis, Health and Toxicology* **2021**, *36*, 1.

[2] S. Umoren, U. Etim, A. Israel, *J. Mater. Environ. Sci* **2013**, *4*, 1 75.

[3] F. Masihi, F. Rezaeitavabe, A. Karimi-Jashni, G. Riefler, *International Journal of Hydrogen Energy* **2023**.

[4] G. Mele, R. Del Sole, X. Lü, In F. Parrino, L. Palmisano, editors, *Titanium Dioxide (Tio2) and Its Applications*, Metal Oxides, 527–581. Elsevier, ISBN 978-0-12-819960-2, **2021**,

[5] U. Chinonso, O. Ibukun, H. Kyung Jeong, *Chemical Physics Letters* **2020**, *757* 137850.

[6] O. Ibukun, P. E. Evans, P. A. Dowben, H. K. Jeong, *Chemical Physics* **2019**, *525* 110419.

[7] G. Kresse, J. Furthmuller, *Phys. Rev. B* **1996**, *54* 11169.

[8] G. Ghanashyam, H. K. Jeong, *Chemical Physics Letters* **2022**, *794* 139492.

[9] G. Ghanashyam, H. K. Jeong, *Journal of Energy Storage* **2021**, *40* 102806.

[10] T. Niyitanga, H. K. Jeong, *Materials Chemistry and Physics* **2021**, *263* 124345.

[11] H. J. Kim, C.-S. Yang, H. Jeong, *Chemical Physics Letters* **2016**, *644* 288.

[12] D. A. Drabold, *Eur. Phys. J. B* **2009**, *68* 1.

[13] B. Prasai, B. Cai, M. K. Underwood, J. P. Lewis, D. Drabold, *Journal of materials science* **2012**, *47*, 21 7515.

[14] B. Prasai, B. Cai, D. Drabold, M. K. Underwood, J. P. Lewis, In *MS\&T-11 Conference of Proceedings*. **2011.**

[15] J. P. Perdew, K. Burke, M. Ernzerhof, *Phys. Rev. Lett.* **1996**, *77* 3865.

[16] A. Monshi, M. R. Foroughi, M. R. Monshi, et al., *World journal of nano science and engineering* **2012**, *2*, 3 154.

[17] A. Patterso, *J. Phys. Rev* **1939**, *56*, 10 978.

[18] L.-Q. Wang, D. R. Baer, M. H. Engelhard, *Surface science* **1994**, *320*, 3 295.

[19] S.-D. Mo, W. Y. Ching, *Phys. Rev. B* **1995**, *51* 13023.

[20] T. Zhu, S.-P. Gao, *The Journal of Physical Chemistry C* **2014**, *118*, 21 11385.

[21] V. Etacheri, C. Di Valentin, J. Schneider, D. Bahnemann, S. C. Pillai, *Journal of Photochemistry and Photobiology C: Photochemistry Reviews* **2015**, *25* 1.

[22] R. Thapa, C. Ugwumadu, K. Nepal, D. Drabold, M. Shatnawi, *Journal of Non-Crystalline Solids* **2023**, *601* 121998.

[23] C. Xu, G. Rangaiah, X. Zhao, *Industrial & Engineering Chemistry Research* **2014**, *53*, 38 14641.

[24] K. V. Kumar, K. Porkodi, F. Rocha, *Catalysis Communications* **2008**, *9*, 1 82.

[25] K. Suriye, P. Praserthdam, B. Jongsomjit, *Applied Surface Science* **2007**, *253*, 8 3849.

**Table and Caption**

| Sample | D [nm] | wt. [%] Ti | O |
|---|---|---|---|
| 0h | 35.91 | 61.41 | 38.59 |
| P1h | 34.42 | 62.70 | 37.29 |
| P2h | 30.44 | 65.14 | 34.86 |
| P3h | 29.18 | 65.73 | 34.37 |
| P4h | 26.13 | 67.99 | 32.01 |

Table 1: Average crystallite sizes (D) calculated using Scherrer formula (column II), and the relative weight percentages of titanium and oxygen in the samples, obtained from the energy dispersive spectroscopy analysis (columns III and IV).

**Figures and Caption**

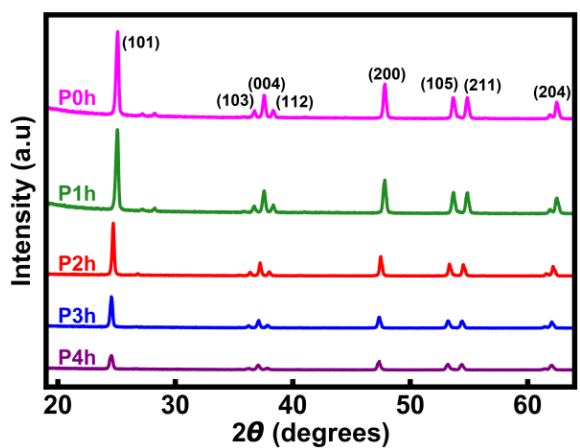

Figure 1: X-ray diffraction spectrum for the samples. The spectra were shifted vertically for clarity.

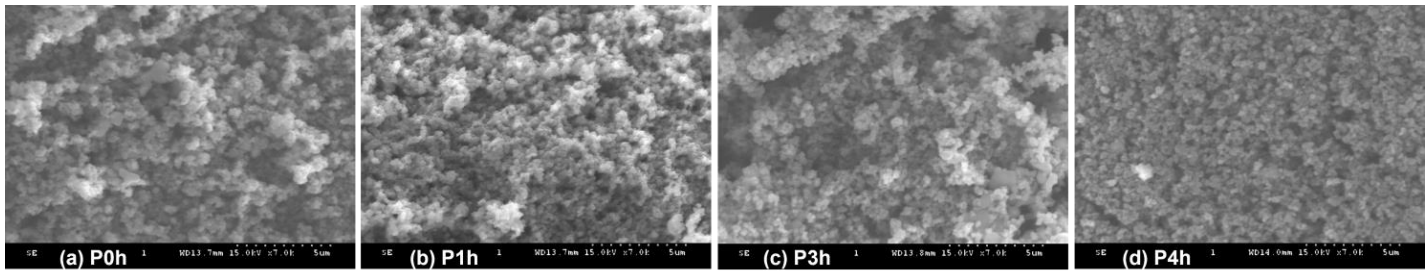

Figure 2: Scanning electron microscope images for (a) P0h, (b) P1h, (c) P3h, and (d) P4h.

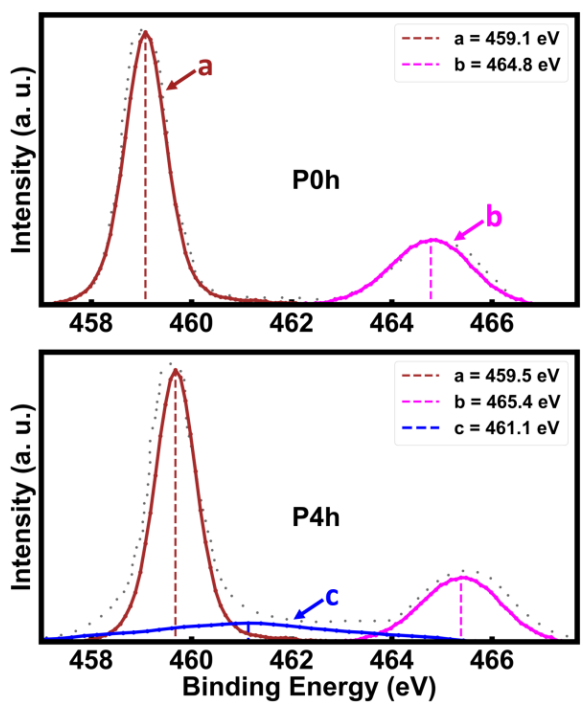

Figure 3: X-ray photoelectron spectroscopy data showing the (a) $Ti^{4+}$ $2p_{3/2}$, (b) $Ti^{4+}$ $2p_{1/2}$, and (c) $Ti^{3+}$ $2p_{3/2}$ states of P0h and P4h samples.

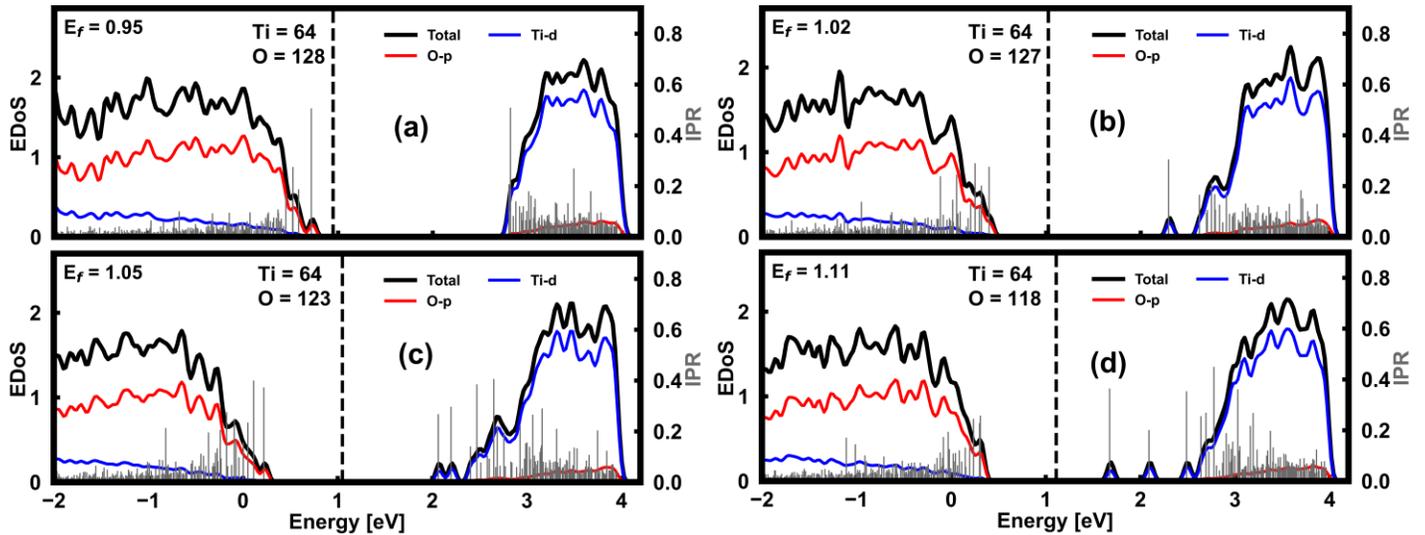

Figure 4: Total (black line) and partial (colored lines) electronic density of states (EDoS) for stoichiometric (a) and non-stoichiometric titanium oxides with 1, 5, and 10 oxygen vacancies (b, c, and d respectively).

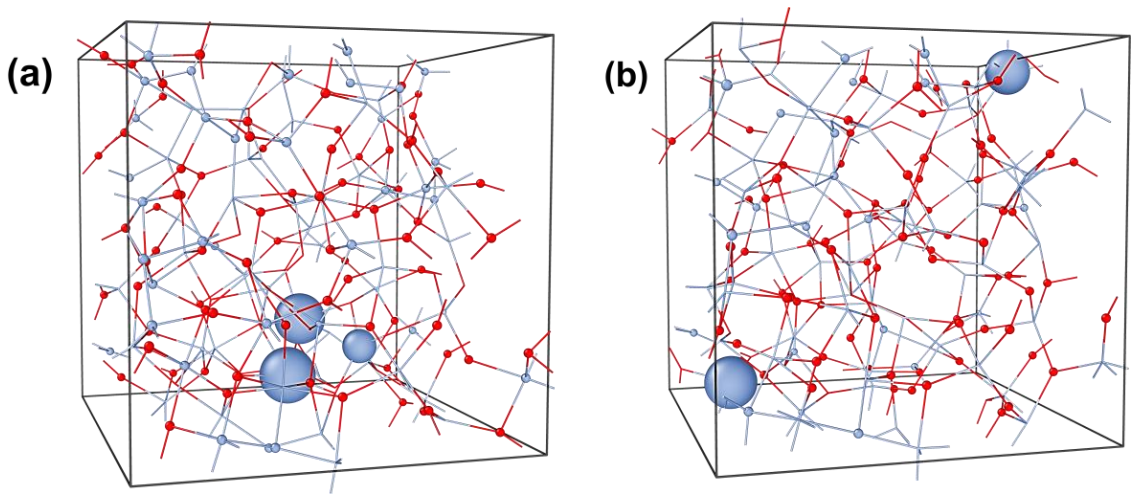

Figure 5: Atoms associated with localized electronic states +2 eV around the Fermi level for non-stoichiometric titanium dioxide with (a) 5 and (b) 10 oxygen-vacancies. The contribution of each atom can be visually gauged by their size; atoms with a large radius are more localized than those having a smaller radius. Titanium (oxygen) atoms are colored blue (red). The gray lines represent the localization of the Kohn-Sham states calculated from the IPR.

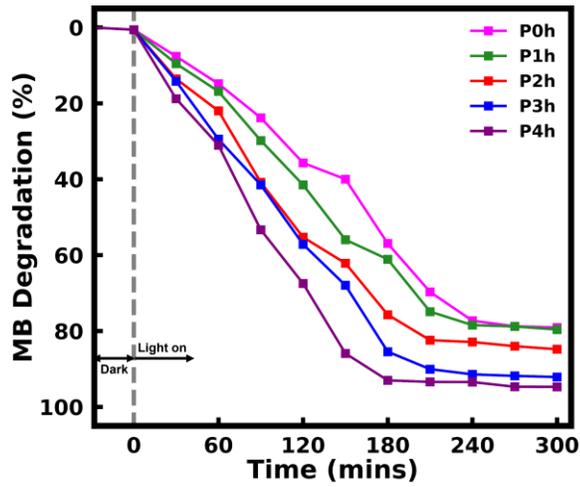

Figure 6: Photocatalytic degradation of MB using the samples.

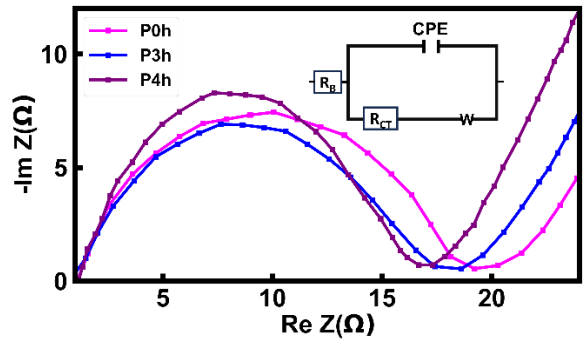

Figure 7: Electrochemical impedance spectroscopy Nyquist plots under ultraviolet irradiation of the samples (Inset is the equivalent circuit).

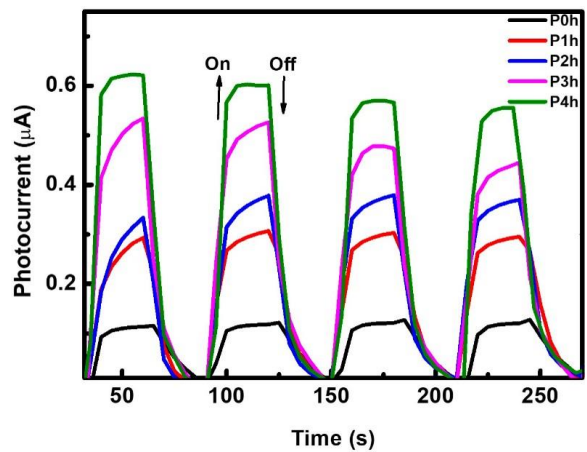

Figure 8: Transient photocurrent responses of the samples under ultraviolet irradiation.

Supplementary Material: Improved Photocatalytic Performance via Air-Plasma Modification of Titanium Dioxide: Insights from Experimental and Simulation Investigation

C. Ugwumadu[1, *], I. Olaniyan[2, 3], H. K. Jeong[4], and D. A. Drabold[1, *]

[1]Department of Physics and Astronomy, Nanoscale and Quantum Phenomena Institute (NQPI), Ohio University, Athens, Ohio 45701, USA.

[2]Helmholtz-Zentrum, Berlin 14109, Germany

[3]Department of Physical Chemistry, Freie Universität Berlin, Berlin 14195, Germany

[4]Department of Materials-Energy Science and Engineering, Institute of Industry and Technique Daegu University, Gyeongsan 38453. Republic of Korea

*Corresponding authors E-mail: cu884120@ohio.edu; drabold@ohio.edu

Abstract:

Commercial titanium dioxide is successfully plasma-treated under ambient conditions for different time periods, leading to reduced crystallite size and the creation of oxygen vacancies. Density functional theory-based calculations reveal the emergence of additional localized states close to the conduction band, primarily associated with under-coordinated titanium atoms in non-stoichiometric titanium-oxide systems. The plasma-treated samples exhibit improved photocatalytic performance in the degradation of methylene blue compared to untreated samples. Moreover, the 4-hour plasma-treated photocatalyst demonstrates commendable stability and reusability. This work highlights the potential of cost-effective plasma treatment as a simple modification technique to significantly enhance the photocatalytic capabilities of titanium dioxide.

Keywords: *titanium dioxide, plasma treatment, photocatalysis, molecular simulation, ultraviolet light*

### Sect. S1. Photo-degradation of Methylene Blue using Air-plasma-treated Titanium-oxide.

Fig.S1a illustrates the degradation process of methylene blue (MB) using the P4h sample over a duration of up to 210 minutes. It provides a visual representation of the gradual degradation of MB over time, showcasing the effectiveness of the P4h sample as a photocatalyst. In Fig.S1b, the UV-Vis spectra are displayed, capturing the absorbance measurements at various wavelengths of light during the degradation of MB using the P4h sample under UV light. The focus is on the most intense peak around 600 nm, which corresponds to a specific absorption characteristic of methylene blue. The change in absorbance at this peak indicates the progress of MB degradation and assesses the efficiency of the photocatalytic process. The close similarity between the UV-Vis spectra obtained at 180 and 210 minutes suggests that the optimal degradation time for the P4h sample is 180 minutes.

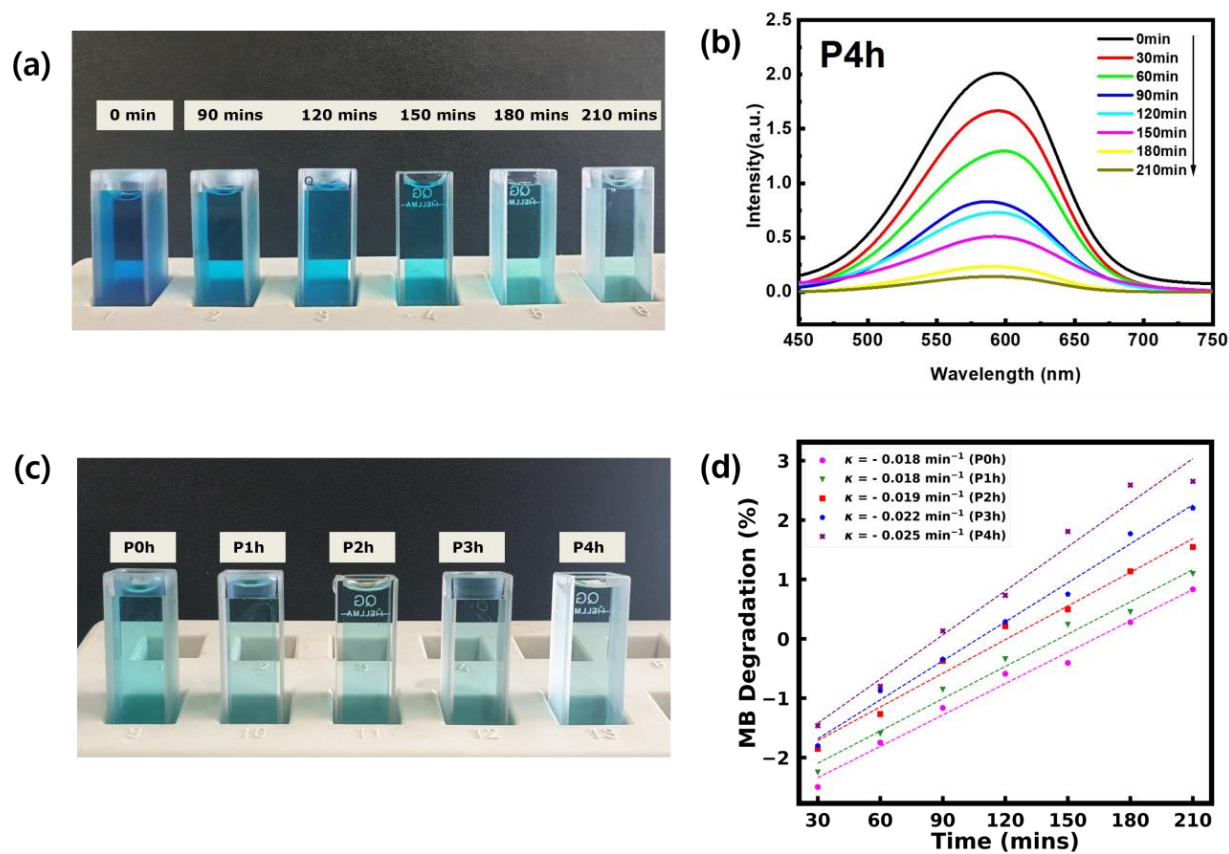

Figure S1: (a) Time-dependent degradation of methylene blue (MB) using the P4h sample over a duration of 210 minutes. (b) UV-Vis spectra showing the absorption peak around 600 nm during MB degradation using P4h under UV light. (c) The resulting solution after complete MB degradation using all the samples. (d) Plot illustrating the degradation rate and kinetic constants.

Fig.S1c presents the final solution obtained after the optimal degradation process using all the samples. This solution represents the product of the degradation process and provides valuable insights into the effectiveness of the photocatalytic activity of the different samples tested. Analyzing the composition and properties of the final solution helps assess the extent of MB degradation and identify any byproducts or residues that may be present. Fig.S1d displays the degradation rate plot, used to evaluate the photocatalytic activity of the samples. It illustrates the apparent kinetic constants associated with the degradation of methylene blue. This plot demonstrates the rate at which MB is being degraded over time for each sample. The reaction rate constants (k) were determined using a first-order approximation of the Langmuir-Hinshelwood equation:

$$-ln\left(\frac{C_0}{C_t}\right) = kKt \approx \kappa t \qquad (1)$$

Here, k and K are the degradation rate constant and equilibrium constant for the adsorption of MB by the catalyst respectively. κ is the apparent rate constant.

### Sect. S2.    Enhanced Hydrophilicity of Air-plasma-treated Titanium-oxide

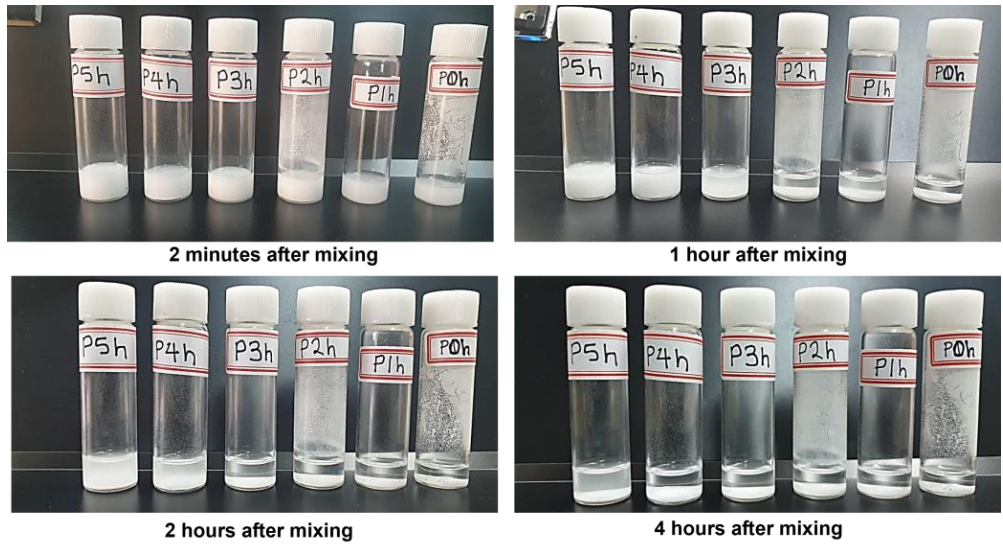

Figure S1: Visual inspection of the samples' hydrophilic characteristics over time

FigureS2 illustrates the hydrophilicity of the air-plasma-treated sample at various time intervals. The samples were mixed with distilled water to create suspension and then allowed to settle on a flat surface. The inclusion of a 5-hour treated sample demonstrates that a treatment duration of 4 hours yields optimal results. The untreated P0h sample settled after just 1 hour, whereas the P4h sample required up to 4 hours to fully settle. These findings indicate that plasma treatment enhanced the hydrophilicity of the titanium-dioxide, resulting in a stronger affinity for water molecules. Consequently, providing more reactive sites for photocatalytic reactions and facilitating the generation of highly reactive hydroxyl radicals (•OH) - crucial for pollutant degradation. Moreover, hydrophilicity of the photocatalyst surface minimizes aggregation or agglomeration, which can impede photocatalytic activity.